\documentclass[preprint,pre,a4paper,11pt,tightenlines,notitlepage,superscriptaddress,longbibliography]{revtex4-1}
\usepackage{graphicx,color,epsfig,subfig,dcolumn,bm,mathrsfs,amsmath}
\usepackage[colorlinks]{hyperref}
\usepackage[T1]{fontenc}

\newcommand{\ud}{\mathrm{d}}

\newcommand{\Tr}{\mathrm{Tr}}

\DeclareMathAlphabet{\mathsfit}{\encodingdefault}{\sfdefault}{m}{sl}
\SetMathAlphabet{\mathsfit}{bold}{\encodingdefault}{\sfdefault}{bx}{sl}
\newcommand{\tens}[1]{\bm{\mathsfit{#1}}}

\newcommand{\bvs}{\mathbf{v}^{(s)}}
\newcommand{\bvp}{\mathbf{v}^{(p)}}

\newcommand{\vp}{{v}^{(p)}}

\newcommand{\bsp}{\tens{\sigma}^{(p)}}

\newcommand{\change}[1]{\textcolor{black}{#1}}

\begin{document}

\title{Derivation of Two-fluid Model Based on Onsager Principle}    

\author{Jiajia Zhou}
\affiliation{South China Advanced Institute for Soft Matter Science and Technology, School of Emergent Soft Matter, South China University of Technology, Guangzhou 510640, China}
\affiliation{Guangdong Provincial Key Laboratory of Functional and Intelligent Hybrid Materials and Devices, South China University of Technology, Guangzhou 510640, China}

\author{Masao Doi}
\affiliation{Center of Soft Matter Physics and its Applications, Beihang University, Beijing 100191, China}
\affiliation{Wenzhou Institute, University of Chinese Academy of Science,
Wenzhou, Zhejiang, 325000, China}


\begin{abstract}
Using Onsager variational principle, we study the dynamic coupling between the stress and the composition in polymer solution. In the original derivation of the two-fluid model [Doi and Onuki, J. Phys. II France {\bf 2}, 1631 (1992)], the polymer stress was introduced \emph{a priopri}, therefore a constitutive equation is required to close the equations. Based on our previous study of viscoelastic fluids \change{with} homogeneous composition [Phys. Rev. Fluids {\bf 3}, 084004 (2018)], we start with a dumbbell model for the polymer, and derive all dynamic equations using the Onsager variational principle. 
\end{abstract}

\maketitle

\newpage

\section{Introduction}

In the \change{studies} of flow for polymeric liquids \cite{BAH1, BAH2}, the inclusion of \change{polymers introduces} two new variables in the system: the polymer concentration and the polymer conformation, which are absent \change{in flow of} simple liquids. 
It was well recognized that time evolution of the microscopic state variable, i.e., the local conformation of the polymer chain, is critically important in governing the dynamics of the polymer solution \cite{Beris_Edwards, Oettinger_beyond}.  
The total stress of a polymer solution is therefore has two contributions, one from the polymer and another from the solvent. 
A prescribed constitutive equation is required to relate the polymer stress to the local flow conditions. 
On the other hand, in the standard treatment for the flow of polymeric liquids \cite{BAH1}, the polymer concentration is assumed to be \change{uniform} in space.
Therefore, the polymer concentration appears as a parameter in the framework, and there is no time evolution equation for the polymer concentration. 

Experimentally, it has been shown that the polymer concentration can become non-uniform when the velocity gradient is not uniform \cite{Wu1991, vanEgmond1992, vanEgmond1993}. 
Theoretically, phenomenological two-fluid model has been developed \cite{Milner1991a, Doi1990, Doi1992} which incorporates the coupling between polymer stress and polymer diffusion in the continuum framework. 
A simple Hookean dumbbell model is used for the polymer chain, and there are a few studies based on different strategies \cite{Bhave1991, Mavrantzas1992, Oettinger1992, Beris1994}. 

In this manuscript, we shall re-derive the two-fluid model based on Onsager principle \cite{DoiSoft}. 
In section \ref{sec:general}, we present a general derivation including all \change{viscous} coupling in the dissipation. 
In section \ref{sec:Doi-Onuki}, we repeat Doi-Onuki's original derivation \change{from} Ref. \cite{Doi1992} for reference. 
In section \ref{sec:dumbbell}, we start with a dumbbell model for the polymer, and derive the time evolution equations using the Onsager variational principle. 
We conclude with a summary in section \ref{sec:conclusion}.












\section{Onsager principle}
\label{sec:general}

\change{Firstly} proposed by Onsager in his celebrated \change{papers} on the reciprocal relation \cite{Onsager1931, Onsager1931a}, Onsager principle is a variational principle to systematically derive the time-evolution equations for \change{out-of-equilibrium systems}.   
The first step is to identify a set of \change{state variables,} $x=(x_1, x_2, \cdots)$, which characterizes the nonequilibrium state of the system under study. 
Then the time evolution of the system is determined by the condition that the following quadratic function of $\dot{x}=(\dot{x}_1, \dot{x}_2, \cdots)$ to be minimized with respect to $\dot{x}$,
\begin{equation}
  \mathscr{R} = \sum_i \frac{\partial A}{\partial x_i} \dot{x}_i
              + \frac{1}{2}  \sum_{i,\,j} \zeta_{ij} \dot{x}_i \dot{x}_j \, . 
  \label{eq:D1}
\end{equation}
Here we use the dot for the partial \change{time derivative}, $\dot{x} = \partial x/\partial t$.

Equation (\ref{eq:D1}) defines the Rayleighian of the system.
\change{It} consists of two parts:
One is the time derivative of the free energy $\dot{A}(x) = \sum_i (\partial A / \partial x_i) \dot{x}_i$. 
The other part is called the dissipation function $\Phi = \frac{1}{2} \sum_{i,j} \zeta_{ij} \dot{x}_i \dot{x}_j$, where $2\Phi$ 
represents the energy dissipated in the system per unit time when the state variables are changing at rate $\dot x$.
The coefficient $\zeta_{ij}$ is called friction coefficient, which is generally a function of state variables \change{$x_i$}.
The dissipation function must be a quadratic function of \change{$\dot{x}_i$}.
The minimum condition of the Rayleighian $\partial \mathscr{R}/\partial  \dot{x}_i =0$ \change{determines} the time evolution of the state variables:
\begin{equation}
  - \frac{\partial A}{\partial x_i} = \sum_j \zeta_{ij} \dot{x}_j  \, .
  \label{eq:D2}
\end{equation}
Equation (\ref{eq:D2}) is an analogue to the force balance equation, where the left-hand side is the 
thermodynamic driving force and the right-hand side is the friction force. 
The reciprocal relation $\zeta_{ij}=\zeta_{ji}$ is required in this derivation.

Onsager principle is particularly useful for soft matter systems when inertia is not important.
Many time evolution equations used in soft matter, such as Stokes equation, Fick's diffusion equation, Nernst-Planck equation, Cahn-Hilliard equation, Ericksen-Leslie equation, etc., can be derived based on Onsager principle \cite{DoiSoft, Doi2012, Doi2021}.
In a previous work \cite{2018_ve_filament}, we have shown that the continuum mechanical equation 
for viscoelastic fluids can also be derived from the Onsager principle.
Here we use the same framework to derive the time evolution equations of two-fluid model \change{for} polymer solutions, by taking into consideration of the coupling between stress and diffusion.

\subsection{State variables}

We first need to identify the state variables that characterize the non-equilibrium state of flowing 
polymer solutions. We choose the state variables as follows:
\begin{itemize}
\item Volume fraction of the polymer $\phi$.

The corresponding ``velocity'' variable is $\bvp$, the polymer velocity. 
The polymer volume fraction and the polymer velocity are related by the conservation law
\begin{equation}
  \label{eq:dot_phi}
  \dot{\phi} = - \bm{\nabla} \cdot \Big( \phi \bvp \Big)
  = - \nabla_{\alpha} \Big( \phi \vp_{\alpha} \Big).
\end{equation}
Here $\dot{\phi} = \partial \phi / \partial t$ and $\nabla_{\alpha} = \partial / \partial x_{\alpha}$.

\item Conformation tensor $\tens{c}$.

\change{$\tens{c}$ is a non-dimensional tensor to characterize the microscopic state of the polymer chain.}
The $\tens{c}$-tensor is equal to unit tensor $\tens{I}$ when the polymer is at 
equilibrium, and deviates from $\tens{I}$ when the polymer is deformed.
Late we will introduce the dumbbell model that presents the polymer chain as a dumbbell consisting of two beads at positions $\mathbf{r}_1$ and $\mathbf{r}_2$. 
These two beads are connected by an elastic spring that has an end-to-end vector $\mathbf{r} = \mathbf{r}_1 - \mathbf{r}_2$ and a spring constant $k$. 
The conformation of the dumbbell is then specified by the $\tens{c}$-tensor defined by
$\tens{c}= \frac{k}{k_BT} \langle \mathbf{r}\mathbf{r} \rangle$.

The corresponding ``velocity'' variable is the material time derivative of $\tens{c}$ defined by
\begin{equation}
  D_t^{(p)} \tens{c} = \frac{\partial \tens{c}}{\partial t} + \bvp \cdot \bm{\nabla} \tens{c}, \quad {\rm or} \,\,\,
  D_t^{(p)} c_{\alpha\beta} = \frac{\partial c_{\alpha\beta} }{\partial t} + \vp_{\gamma} \nabla_{\gamma} c_{\alpha\beta} .
\end{equation}
Notice that here we are using the polymer velocity $\bvp$ to define the material time derivative, not the 
medium velocity $\mathbf{v}$  which will be introduced next.  

\item In order to discuss the phenomena of diffusion or migration of polymers,  
we need to introduce another ``velocity'' variable  representing the velocity of the surrounding.  This
can be represented by the solvent velocity $\bvs$, or the medium velocity (volume-average velocity) 
defined by
\begin{equation}
  \label{eq:v_vol_ave}
  \mathbf{v} = \phi \bvp + (1-\phi) \bvs .
\end{equation}
Here we will use $\mathbf{v}$ following rheology convention. 
\end{itemize}

The local flow condition is characterized by the velocity gradient tensor $\bm{\nabla} \mathbf{v}$
\begin{equation} 
  \nabla_{\alpha} v_{\beta} = \frac{\partial v_{\beta}} {\partial x_{\alpha}},
\end{equation}
and the related rate-of-strain tensor
\begin{equation}
  \dot{\tens{\gamma}} = (\bm{\nabla} \mathbf{v})^t + \bm{\nabla} \mathbf{v} 
  = \left( \frac{\partial v_{\alpha}}{\partial x_{\beta}} 
    + \frac{\partial v_{\beta}}{\partial x_{\alpha}} \right) .
\end{equation}

\subsection{Free energy}

The general form of the free energy of polymer solution can be written as 
\begin{equation}
  A = \int \ud \mathbf{r} \, a( \phi, \tens{c}).
\end{equation}
where $a( \phi, \tens{c})$ is the free energy density. 
We assume that $a( \phi, \tens{c})$ has the following form
\begin{equation}
  a( \phi, \tens{c}) = f(\phi) + \phi g(\tens{c}).
\end{equation}
The first term $f(\phi)$ is the free energy density of polymer solutions at equilibrium. 
This term includes the entropic term $\phi \ln\phi$, \change{the interaction term of Flory-Huggins form} $\chi \phi(1-\phi)$, and the interfacial energy which depends on 
the concentration gradient $|\nabla \phi|^2$. 
\change{The second term includes $g(\tens{c})$ that represents the elastic energy of deformed polymer chains and is a function of the conformation tensor $\tens{c}$.
The second term is proportional to the polymer volume fraction $\phi$.}

The change rate of the free energy is given by
\begin{eqnarray}
  \dot{A} &=& \int \ud \mathbf{r} \Bigg[ \frac{\partial a}{\partial \phi} \dot{\phi}
    + \frac{\partial a}{\partial c_{\alpha \beta}} \dot{c}_{\alpha \beta} \Bigg] \nonumber \\
  &=& \int \ud \mathbf{r} \Bigg[ - \frac{\partial a}{\partial \phi} 
  \nabla_{\gamma} \big( \phi \vp_{\gamma} \big) 
  + \frac{\partial a}{\partial c_{\alpha \beta}} \dot{c}_{\alpha \beta} \Bigg] \nonumber \\
&=& \int \ud \mathbf{r} \Bigg[ \phi \vp_{\gamma} 
    \nabla_{\gamma} \big( \frac{\partial a}{\partial \phi} \big) 
  + \frac{\partial a}{\partial c_{\alpha \beta}} 
  \Big(D_t^{(p)}{c}_{\alpha \beta} - \vp_{\gamma} \nabla_{\gamma} c_{\alpha \beta} \Big) \Bigg] \nonumber \\   
&=& \int \ud \mathbf{r} \Bigg[ 
    \vp_{\gamma} \nabla_{\gamma} \big( \phi \frac{\partial a}{\partial \phi} \big)
    - \vp_{\gamma} \frac{\partial a}{\partial \phi} \nabla_{\gamma} \phi
    - \vp_{\gamma} \frac{\partial a}{\partial c_{\alpha\beta}} \nabla_{\gamma} c_{\alpha\beta}
    + \frac{\partial a}{\partial c_{\alpha\beta}} D_t^{(p)} c_{\alpha\beta} \Bigg] \nonumber \\
&=& \int \ud \mathbf{r} \Bigg[ 
    \vp_{\gamma} \nabla_{\gamma} \big( \phi \frac{\partial a}{\partial \phi} \big)
    - \vp_{\gamma} \nabla_{\gamma} a
    + \frac{\partial a}{\partial c_{\alpha\beta}} D_t^{(p)} c_{\alpha\beta} \Bigg] \nonumber \\
&=& \int \ud \mathbf{r} \Bigg[ 
    \vp_{\gamma} \nabla_{\gamma} \big( \phi \frac{\partial a}{\partial \phi} -a  \big)
    + \frac{\partial a}{\partial c_{\alpha\beta}} D_t^{(p)} c_{\alpha\beta} \Bigg] \nonumber \\
&=& \int \ud \mathbf{r} \Bigg[ 
    \vp_{\gamma} \nabla_{\gamma} \Pi
    + \frac{\partial a}{\partial c_{\alpha\beta}} D_t^{(p)} c_{\alpha\beta} \Bigg] \,.
    \label{eq:Adot_2f}
\end{eqnarray}
where $\Pi$ stands for the osmotic pressure and is \change{defined} by 
\begin{equation}
  \label{eq:Pi}
  \Pi \equiv \phi \frac{\partial a}{\partial \phi} -a .
\end{equation}
Notice that the elastic energy term $g(\tens{c})$ has no contribution to 
the osmotic pressure
\begin{equation}
  \Pi = \phi \frac{\partial (f+\phi g)}{\partial \phi} - (f+\phi g) 
      = \phi \frac{\partial f}{\partial \phi} - f .
\end{equation}

\subsection{Dissipation function}

The dissipation function also includes several terms. 
The first one accounts the relative motion of the center-of-mass of the polymer with respect to the 
medium velocity
\begin{equation}
  \label{eq:Phi_pv}
  \Phi_{pv} = \frac{1}{2} \int \ud\mathbf{r} \, \xi (\bvp - \mathbf{v})^2,
\end{equation}
where the friction coefficient $\xi=\xi(\phi)$ is in general concentration-dependent. 
Notice that by use of Eq. (\ref{eq:v_vol_ave}), the integrand of Eq. (\ref{eq:Phi_pv}) is written as $\xi (1- \phi)^2 (\bvp - \bvs)^2$. Therefor $\Phi_{pv}$  can be understood as the dissipation due to the relative motion between polymer and solvent.

The second term represents the coupling between $D_t^{(p)} \tens{c}$ and $\bm{\nabla} \mathbf{v}$. 
We write it in a very general form
\begin{equation}
  \label{eq:Phi_cv}
  \Phi_{cv} = \frac{1}{2} \int \ud\mathbf{r} \left\{ 
    \xi^{(cc)}_{\alpha\beta\mu\nu} (D_t^{(p)} c_{\alpha\beta})(D_t^{(p)} c_{\mu\nu})  
    + 2 \xi^{(cv)}_{\alpha\beta\mu\nu} (D_t^{(p)} c_{\alpha\beta}) (\nabla_{\mu} v_{\nu}) 
    + \xi^{(vv)}_{\alpha\beta\mu\nu} (\nabla_{\beta} v_{\alpha}) (\nabla_{\mu} v_{\nu}) 
    \right\}
\end{equation}
The first term is the inter-coupling of $D_t^{(p)} \tens{c}$, the second term is the cross-coupling between $D_t^{(p)} \tens{c}$ and $\bm{\nabla} \mathbf{v}$, and the last term is related to the solvent viscosity.
Since $D_t^{(p)} \tens{c}$ and $\bm{\nabla} \mathbf{v}$ are tensors of rank 2, the frictional coefficients $\zeta$ are tensors of rank 4.
These coefficients must be positive-definite to ensure that the dissipation function is non-negative.

\subsection{Time evolution equations}

\change{From the change rate of the free energy (\ref{eq:Adot_2f}) and the dissipation functions (\ref{eq:Phi_pv}) and (\ref{eq:Phi_cv}), the Rayleighian can be written as}
\begin{eqnarray}
  \mathscr{R} &=& \int \ud \mathbf{r} \Bigg[ \vp_{\alpha} \nabla_{\alpha} \Pi
  + \frac{\partial a}{\partial c_{\alpha \beta}} D_t^{(p)}{c}_{\alpha \beta} \Bigg] 
  + \frac{1}{2} \int \ud\mathbf{r} \xi (\vp_{\alpha} - v_{\alpha})^2 \nonumber \\
 && + \frac{1}{2} \int \ud\mathbf{r} \left\{ 
    \xi^{(cc)}_{\alpha\beta\mu\nu} (D_t^{(p)} c_{\alpha\beta})(D_t^{(p)} c_{\mu\nu}) 
    + 2 \xi^{(cv)}_{\alpha\beta\mu\nu} (D_t^{(p)} c_{\alpha\beta}) (\nabla_{\mu} v_{\nu}) 
    +  \xi^{(vv)}_{\alpha\beta\mu\nu} (\nabla_{\beta} v_{\alpha}) (\nabla_{\mu} v_{\nu})
    \right\} \nonumber \\
 \label{eq:Ray}
 && - \int \ud\mathbf{r} \, p (\nabla_{\alpha} v_{\alpha}),
\end{eqnarray}
where the last term accounts for the incompressibility condition $\bm{\nabla} \cdot \mathbf{v}=0$.

By the variational calculation with respect to the three ``velocity'' variables, we get the following set of  equations 
\begin{eqnarray}
  \label{eq:dRdvp}
  \frac{ \delta \mathscr{R} }{\delta \vp_{\alpha}} = 0 & \,\, \Rightarrow \,\, &
    \xi ( \vp_{\alpha} - v_{\alpha} )
    + \nabla_{\alpha} \Pi = 0 \\
  \label{eq:dRdva}
  \frac{ \delta \mathscr{R} }{\delta v_{\alpha}} = 0 & \,\, \Rightarrow \,\, &
    - \xi ( \vp_{\alpha} - v_{\alpha} ) 
    - \nabla_{\beta} \Big( \xi^{(cv)}_{\mu\nu\alpha\beta} (D_t^{(p)} c_{\mu\nu}) \Big)
    - \nabla_{\beta} \Big( \xi^{(vv)}_{\mu\nu\alpha\beta} (\nabla_{\mu} v_{\nu}) \Big)
    + \nabla_{\alpha} p = 0 \\
  \label{eq:dRdc}
  \frac{ \delta \mathscr{R} }{\delta D_t^{(p)} c_{\alpha\beta}} = 0 
    & \,\, \Rightarrow \,\, &
    \xi^{(cc)}_{\alpha\beta\mu\nu} (D_t^{(p)} c_{\mu\nu}) 
    + \xi^{(cv)}_{\alpha\beta\mu\nu} (\nabla_{\mu} v_{\nu}) 
    + \frac{\partial a}{\partial c_{\alpha\beta} } = 0 
\end{eqnarray}

Equation (\ref{eq:dRdvp}) and the conservation equation (\ref{eq:dot_phi}) lead to the time-evolution equation for $\phi$.
Equations (\ref{eq:dRdvp}) and (\ref{eq:dRdva}) give the following force balance equation
\begin{equation}
    \nabla_{\beta} \Big( - \xi^{(cv)}_{\mu\nu\alpha\beta} (D_t^{(p)} c_{\mu\nu}) 
    - \xi^{(vv)}_{\mu\nu\alpha\beta} (\nabla_{\mu} v_{\nu}) 
    + ( \Pi + p ) \delta_{\alpha \beta} \Big) = 0 \, ,
\end{equation}
from which we can identify the stress tensor as the terms in the big brackets.
Equation (\ref{eq:dRdc}) gives the constitutive equation
\begin{equation}
  \label{eq:Dtp}
    D_t^{(p)} c_{\mu\nu} = 
    - ( \nabla_{\gamma} v_{\epsilon}) \xi^{(cv)}_{\alpha\beta\gamma\epsilon} \xi^{(cc)^{-1}}_{\alpha\beta\mu\nu}
    - \frac{\partial a}{\partial c_{\alpha\beta} } \xi^{(cc)^{-1}}_{\alpha\beta\mu\nu} .
\end{equation}

These results are quite general, but their usefulness is limited because many phenomenological parameters are
introduced in the model.  It is not clear how to assign values to these parameters for a practical polymer model. 
In the following we shall study two existing theories from the view point of this general formulation.


\section{Doi-Onuki derivation}
\label{sec:Doi-Onuki}

\subsection{Polymer solution}

Doi and Onuki \cite{Doi1992}, and Mavrantzas and Beris \cite{Mavrantzas1992}
proposed \change{two-fluid} model for polymer solutions.  
Their theories are not equivalent to each other, but  are based on the same idea that the 
elastic stress created in the polymer  will contribute to the motion of the polymer 
relative to the solvent if the stress is not uniform.  
Here we will discuss \change{the two-fluid model} focusing on the Doi-Onuki theory.  
They started with a free energy change rate of the form
\begin{equation}
  \dot{F} = \dot{F}_{\rm mix} + \dot{F}_{\rm el} .
\end{equation}
The first term $\dot{F}_{\rm mix}$ is the change rate of the mixing free energy 
\begin{eqnarray}
  F_{\rm mix} &=& \int \ud \mathbf{r} f(\phi), \\
  \dot{F}_{\rm mix} &=& \int \ud \mathbf{r} \frac{\partial f}{\partial \phi} \dot{\phi} 
    = - \int \ud \mathbf{r} \frac{\partial f}{\partial \phi} \bm{\nabla} \cdot (\phi \bvp)
    = \int \ud \mathbf{r} \bvp \cdot \bm{\nabla} \Pi
\end{eqnarray}
with $\Pi$ is the osmotic pressure \change{given by} Eq.~(\ref{eq:Pi}).
The second term $\dot{F}_{\rm el}$ is the change rate of the elastic free energy (Eq. (2.26) of Ref. \cite{Doi1992})
and defines the stress exerted on the polymer $\bsp$ (called the ``network stress'' $\bm{\sigma}^{(n)}$ in Ref. \cite{Doi1992}).
\begin{equation}
  \label{eq:dFel}
  \dot{F}_{\rm el} = \int \ud \mathbf{r} \bsp : (\bm{\nabla} \bvp).
\end{equation}

Some comments on this term are in order:
\begin{itemize}
\item This derivation is different to the standard derivation based on Onsager principle. 
We normally start with a free energy as a function of the state variables and then calculate the change rate by performing the time derivative. 
\item Here the polymer stress $\bsp$ is put in by hand, therefore we still need a constitutive equation to relate the stress to the state variables. 
\item Here the polymer velocity is used. In Ref. \cite{Doi1992}, it was noted ``\emph{since it is the deformation of the polymer which causes the change of the free energy.}'' It turns out it is very important to specify which velocity is coupled to the polymer stress. 
\end{itemize}

The dissipation function is given by 
\begin{equation}
  \Phi = \frac{1}{2} \int \ud \mathbf{r}\, \zeta (\bvp - \bvs)^2 ,
\end{equation}
due to the relative motion of the polymer chain to the solvent.

The Rayleighian is 
\begin{equation}
  \mathscr{R} = \int \ud \mathbf{r} \left[ \frac{1}{2} \zeta (\bvp - \bvs)^2 
     + \bvp \cdot \bm{\nabla} \Pi 
     + \bsp : \bm{\nabla} \bvp 
     - p \bm{\nabla} \cdot( \phi \bvp + (1-\phi) \bvs) \right]
\end{equation}
which gives
\begin{eqnarray}
  \label{eq:DO_ps1}
  \frac{\delta \mathscr{R}}{\delta \bvp} = 0 & \quad \Rightarrow \quad &
     \zeta (\bvp - \bvs) - \bm{\nabla} \cdot \bsp + \bm{\nabla} \Pi + \phi \bm{\nabla} p = 0 ,\\
  \label{eq:DO_ps2}
  \frac{\delta \mathscr{R}}{\delta \bvs} = 0 & \quad \Rightarrow \quad &
     \zeta (\bvs - \bvp) + (1-\phi) \bm{\nabla} p = 0 .
\end{eqnarray}
These are Eqs.~(2.28) and (2.29) in Ref.~\cite{Doi1992}. 
Here we have only two time-evolution equations.  
To complete the formulation, a constitutive equation needs to be specified. 
This is slightly different from the previous derivation, where the constitutive equation (\ref{eq:Dtp}) is derived from the condition $\delta \mathscr{R}/ \delta D_t^{(p)} c_{\alpha\beta} = 0$. 

\subsection{Polymer blends}

For binary mixture of the polymer melt, Doi and Onuki suggested to use the ``tube velocity'' (or friction-average velocity, also see Tanaka's works \cite{Tanaka1997a, Tanaka2000b}) in Eq.~(\ref{eq:dFel})
\begin{equation}
  \mathbf{v}_{\rm T} = \frac{1}{\zeta_{\rm L} + \zeta_{\rm S}} ( \zeta_{\rm L} \mathbf{v}_{\rm L} + \zeta_{\rm S} \mathbf{v}_{\rm S} )
\end{equation}
where L and S stand for the long and short polymers. 

The Rayleighian is written as
\begin{eqnarray}
  \mathscr{R} &=& \int \ud \mathbf{r} \Big[ \frac{1}{2} \zeta (\mathbf{v}_{\rm L} - \mathbf{v}_{\rm S})^2 
     - \mu \bm{\nabla} \cdot(\phi_{\rm L} \mathbf{v}_{\rm L})
     + \bm{\sigma}^{(n)} : \bm{\nabla} (\zeta_{\rm L} + \zeta_{\rm S})^{-1} ( \zeta_{\rm L} \mathbf{v}_{\rm L} + \zeta_{\rm S} \mathbf{v}_{\rm S} )  \nonumber \\
     && - p \bm{\nabla} \cdot( \phi_{\rm L} \mathbf{v}_{\rm L} + (1-\phi_{\rm L}) \mathbf{v}_{\rm S}) \Big]
\end{eqnarray}
which gives
\begin{eqnarray}
  \label{eq:DO_pb1}
  \frac{\delta \mathscr{R}}{\delta \mathbf{v}_{\rm L}} = 0 & \quad \Rightarrow \quad &
     \zeta (\mathbf{v}_{\rm L} - \mathbf{v}_{\rm S}) 
     - \frac{\zeta_{\rm L}}{\zeta_{\rm L} + \zeta_{\rm S}} \bm{\nabla} \cdot \bm{\sigma}^{(n)} 
     + \phi_{\rm L} \bm{\nabla} \mu  + \phi_{\rm L} \bm{\nabla} p = 0 ,\\
  \label{eq:DO_pb2}
  \frac{\delta \mathscr{R}}{\delta \mathbf{v}_{\rm S}} = 0 & \quad \Rightarrow \quad &
     \zeta (\mathbf{v}_{\rm S} - \mathbf{v}_{\rm L}) 
     - \frac{\zeta_{\rm S}}{\zeta_{\rm L} + \zeta_{\rm S}} \bm{\nabla} \cdot \bm{\sigma}^{(n)} 
     + (1-\phi_{\rm L}) \bm{\nabla} p = 0 .
\end{eqnarray}
These are Eqs.~(4.3) and (4.4) of Ref.~\cite{Doi1992}.

\section{Dumbbell model}
\label{sec:dumbbell}

A popular model of viscoelastic fluid is Oldroyd-B model, \change{also known as} the dumbbell model \cite{BAH1, BAH2}.  
It has been shown \cite{Beris_Edwards}  that this model can be derived from an 
energetic principle similar to the Onsager principle.  
In our previous work \cite{2018_ve_filament} we derived a Rayleighian which gives the Oldroyd-B model, and showed that such formulation is useful to \change{obtain} analytical solutions for certain problems.  
Here we extend the framework to the two-fluid model, and derive a set of equations which 
accounts for the coupling of the flow and the diffusion. 

A polymer chain is modeled as a dumbbell consisting of two beads connected by an elastic 
spring that has an end-to-end vector $\mathbf{r} = \mathbf{r}_1 - \mathbf{r}_2$. 
The conformation of the dumbbell is specified by the $\tens{c}$-tensor defined by
$\tens{c}= \frac{k}{k_BT} \langle \mathbf{r}\mathbf{r} \rangle$. 

\subsection{Free energy}

For dilute solutions, the free energy function is given by
\begin{eqnarray}
  a(\phi, \tens{c}) &= & \frac{k_BT}{v}  \phi \ln \phi + \phi g(\tens{c}) ,  \\
  g(\tens{c}) &=& \frac{1}{2} \frac{k_BT}{v} \left[ \Tr( \tens{c} ) - \ln \det (\tens{c}) \right] ,
\end{eqnarray}
where $v$ is the volume of one single dumbbell.
The change rate of the free energy is given by Eq.~(\ref{eq:Adot_2f})
\begin{equation}
  \dot{A} = \int \ud \mathbf{r} \Bigg[ 
    \vp_{\gamma} \nabla_{\gamma} \Pi
    + \phi \frac{\partial g}{\partial c_{\alpha\beta}} D_t^{(p)} c_{\alpha\beta} \Bigg] \, .
\end{equation}
The variation of $\dot{A}$  with respect to the velocity variables are  
\begin{eqnarray}
  \frac{ \delta \dot{A}}{\delta \bvp} &=& \boldsymbol{\nabla} \Pi \, ,\\
  \frac{ \delta \dot{A}}{\delta \bvs} &=& 0 \, ,\\
  \frac{ \delta \dot{A}}{\delta D_t^{(p)} \tens{c}} 
    &=& \phi \frac{\partial g}{\partial \tens{c}} 
     =  \frac{1}{2} \frac{k_BT}{v} \phi ( \tens{I} - \tens{c}^{-1} ) \, .
\end{eqnarray}

\subsection{Dissipation function}

The dissipation function related to the $\tens{c}$-tensor is given in Ref.~\cite{2018_ve_filament}, 
with $\dot{\tens{c}}$ replaced by $D_t^{(p)}\tens{c}$,
\begin{equation}
  \label{eq:Phic}
  \Phi_c = \frac{1}{4} \frac{k_BT}{v} \int \ud \mathbf{r} \tau \phi {\rm Tr} 
  \left[ \tens{c}^{-1} \cdot 
    ( D_t^{(p)}\tens{c}^t - \tens{\kappa} \cdot \tens{c} - \tens{c} \cdot \tens{\kappa}^t) \cdot
    ( D_t^{(p)}\tens{c}   - \tens{\kappa} \cdot \tens{c} - \tens{c} \cdot \tens{\kappa}^t) \right] \, ,
\end{equation}
where ${\rm Tr}$ denotes the trace operation.

Here comes the main question: 
Which velocity \change{should we use} in the velocity gradient $\tens{\kappa}$? 
We have a few options here:

\begin{itemize}
  \item using the polymer velocity, $\tens{\kappa} = \tens{\kappa}^{(p)} = (\bm{\nabla} \bvp)^t$. This will give the same results of Doi-Onuki.
  \item using the solvent velocity, $\tens{\kappa} = \tens{\kappa}^{(s)} = (\bm{\nabla} \bvs)^t$. This will lead to a different dynamics.
  \item using a combination of the polymer and solvent velocities, $\tens{\kappa} = (\bm{\nabla} \mathbf{V})^t$, with
    \begin{equation}
      \label{eq:bigV}
      \mathbf{V} = \alpha_1 \bvp + \alpha_2 \bvs, \quad \alpha_1 + \alpha_2 = 1.
    \end{equation}
\change{This includes a special case of volume-average velocity $\mathbf{V}=\mathbf{v}$, with $\alpha_1=\phi$ and $\alpha_2=1-\phi$.} 
\end{itemize}

We will continue the derivation using the combination velocity $\mathbf{V}$. 
In general one should expect $\alpha_i$ to be a function of the \change{concentration $\phi$ so it will be position-dependent}. 
\begin{equation}
  \tens{\kappa} = (\bm{\nabla}\mathbf{V})^t = \change{\left[\bm{\nabla}(\alpha_1 \bvp + \alpha_2 \bvs) \right]^t}.
\end{equation}

The total dissipation function is given by
\begin{equation}
  \Phi = \Phi_c + \frac{1}{2}   \int \ud \mathbf{r} \zeta (\bvp - \bvs)^2.
\end{equation}

The variation with respect to the dynamic variables are
\begin{eqnarray}
  \frac{ \delta \Phi}{\delta \bvp} &=& \zeta (\bvp - \bvs) 
     + \bm{\nabla} \cdot \left[ \change{\alpha_1} \frac{k_BT}{v} \tau  \phi \big( D_t^{(p)}\tens{c} 
        - (\bm{\nabla}\mathbf{V})^t \cdot \tens{c} - \tens{c} \cdot (\bm{\nabla}\mathbf{V}) \big) \right]  \, , \\ 
  \frac{ \delta \Phi}{\delta \bvs} 
    &=&  \zeta (\bvs - \bvp) 
     + \bm{\nabla} \cdot \left[ \change{\alpha_2} \frac{k_BT}{v} \tau  \phi \big( D_t^{(p)}\tens{c} 
        - (\bm{\nabla}\mathbf{V})^t \cdot \tens{c} - \tens{c} \cdot (\bm{\nabla}\mathbf{V}) \big) \right]  \, , \\
  \frac{ \delta \Phi}{\delta D_t^{(p)} \tens{c}} 
    &=& \frac{1}{2} \frac{k_BT}{v} \tau \phi \big( D_t^{(p)}\tens{c} - (\bm{\nabla}\mathbf{V})^t \cdot \tens{c} 
        - \tens{c} \cdot (\bm{\nabla}\mathbf{V}) \big)  \cdot \tens{c}^{-1} \, .
\end{eqnarray}

\subsection{Time evolution equations}

The Rayleighian is written as
\begin{equation}
  \mathscr{R} = \dot{A} + \Phi - \int \ud \mathbf{r} p \boldsymbol{\nabla} \cdot (\phi \bvp + (1-\phi) \bvs) .
\end{equation}

The variation of the Rayleighian with respect to $D_t^{(p)}\tens{c}$ gives
\begin{equation}
  \label{eq:constitutive}
  D_t^{(p)}\tens{c} - (\bm{\nabla}\mathbf{V})^t \cdot \tens{c} - \tens{c} \cdot (\bm{\nabla}\mathbf{V})
    = - \frac{1}{\tau} ( \tens{c} - \tens{I} ).
\end{equation}
This is the constitutive equation of the Oldroyd-B fluid model, \change{with a small modification that the velocity gradient is given by $\mathbf{V}$.}

The variation with respect to $\bvp$ gives
\begin{equation}
  \label{eq:dbvp1}
  \zeta (\bvp - \bvs) + \bm{\nabla} \cdot \left[ \change{\alpha_1} \frac{k_BT}{v} \tau  \phi \big( D_t^{(p)}\tens{c} - (\bm{\nabla}\mathbf{V})^t \cdot \tens{c} - \tens{c} \cdot (\bm{\nabla}\mathbf{V}) \big) \right] + \bm{\nabla}\Pi + \phi \bm{\nabla} p = 0.
\end{equation}

The variation with respect to $\bvs$ gives
\begin{equation}
  \label{eq:dbvs1}
  \zeta (\bvs - \bvp) + \bm{\nabla} \cdot \left[ \change{\alpha_2} \frac{k_BT}{v} \tau  \phi \big( D_t^{(p)}\tens{c} - (\bm{\nabla}\mathbf{V})^t \cdot \tens{c} - \tens{c} \cdot (\bm{\nabla}\mathbf{V}) \big) \right]  + (1-\phi) \bm{\nabla} p = 0.
\end{equation}

Equations (\ref{eq:dbvp1}) and (\ref{eq:dbvs1}) give 
\begin{equation}
  \label{eq:sum}
  \bm{\nabla} \cdot \left[ \bsp -  (\Pi + p) \tens{I} \right ] = 0 \, ,
\end{equation}
where $\bsp$ is defined by
\begin{equation}
  \label{eq:sigma_p} 
  \bsp = - \frac{k_BT}{v} \tau  \phi \big( D_t^{(p)}\tens{c} 
  - (\bm{\nabla}\mathbf{V})^t \cdot \tens{c} 
  - \tens{c} \cdot (\bm{\nabla}\mathbf{V}) \big) 
  = \frac{k_BT}{v} \phi (\tens{c} - \tens{I}) .
\end{equation}

Equation (\ref{eq:sum}) indicates that the tensor $\tens{\sigma}^{(p)} -  (\Pi + p) \tens{I}$ is the total stress tensor and $\tens{\sigma}^{(p)}$ is the polymer contribution to the stress tensor.
\change{This definition does not depend on the choice of $\alpha_i$.}
For homogeneous solution, the polymer number density is given by $n_p=\phi/v$, \change{then} Eq.~(\ref{eq:sigma_p}) becomes the standard form $\bsp = n_p k_BT (\tens{c}-\tens{I}) = G(\tens{c}-\tens{I})$ with the shear modulus $G=n_p k_BT$. 

Using the expression (\ref{eq:sigma_p}), Eqs. (\ref{eq:dbvp1}) and (\ref{eq:dbvs1}) can then be written as
\begin{eqnarray}
  \label{eq:dbvp2}
  && \zeta (\bvp - \bvs) - \bm{\nabla} \cdot ( \change{\alpha_1} \bsp ) 
     + \bm{\nabla}\Pi + \phi \bm{\nabla} p = 0 \, , \\
  \label{eq:dbvs2}
  && \zeta (\bvs - \bvp) - \bm{\nabla} \cdot ( \change{\alpha_2} \bsp )  
     + (1-\phi) \bm{\nabla} p = 0 \, . 
\end{eqnarray}

With the setting $\alpha_1=1$ and $\alpha_2=0$, we recover the Doi-Onuki results for the polymer solutions (\ref{eq:DO_ps1}) and (\ref{eq:DO_ps2}). 
With the setting $\alpha_1=\zeta_{\rm L}/(\zeta_{\rm L} + \zeta_{\rm S})$ and $\alpha_2=\zeta_{\rm S}/(\zeta_{\rm L} + \zeta_{\rm S})$, we recover the Doi-Onuki results for the polymer blends (\ref{eq:DO_pb1}) and (\ref{eq:DO_pb2}).

Combining the above two equations, we can obtain 
\begin{equation}
  \label{eq:final}
  \zeta (\bvp - \bvs) = - (1-\phi) \bm{\nabla}\Pi + \change{( \alpha_1 - \phi ) \bm{\nabla} \cdot \bsp + (\bm{\nabla} \alpha_1) \cdot \bsp} ,
\end{equation}
\change{where the last term only appears if $\alpha_1$ is position-dependent.}

\change{The above equation shows that the relative motion of the polymers with respect to the solvent has two origins: 
One is the gradient of the osmotic pressure $\bm{\nabla}\Pi$, which corresponds to the usual diffusion due to the concentration gradient. 
The other one is the gradient of the polymer contribution to the stress tensor $\bm{\nabla} \cdot \bsp$. 
This is the essence of the two-fluid model: the polymer contribution to the stress tensor should induce the polymer diffusion, i.e., the stress and the diffusion are coupled. 
Using the conservation equation, we can see that the time-derivative of the volume fraction has contributions from $\nabla^2 \Pi$ and $\bm{\nabla}\bm{\nabla}:\bsp$.
These two terms are consistent with previous two-fluid models \cite{Beris1994}.
Existing models have used $\alpha_i$ that are independent of the position, therefore the last term in Eq.~(\ref{eq:final}) vanishes.}

\change{The magnitude of the stress contribution to the diffusion depends on the choice of $\alpha_i$. 
For $\alpha_1=1$ and $\alpha_2=0$, the stress contribution is on the order of ${O}(1-\phi)$, the same order of the contribution from the osmotic pressure.
This is the original Doi-Onuki result \cite{Doi1992}. 
For $\alpha_1=0$ and $\alpha_2=1$, the stress contribution is reduced to order ${O}(\phi)$, which is small in comparison to the osmotic pressure term.
For $\alpha_1=\phi$ and $\alpha_2=1-\phi$, the stress contribution vanishes.}

For homogeneous solutions, the polymer concentration is uniform in space, therefore the polymer velocity $\bvp$ and the solvent velocity $\bvs$ are the same, $\bvp = \bvs$. 
Different choices in the velocity gradient $\boldsymbol{\kappa}$ lead to the same time-evolution equations.

\section{Conclusion}
\label{sec:conclusion}

We \change{have} used Onsager variational principle to derive the time-evolution equations for polymer solutions, taking into consideration of the coupling between the stress and the composition. 
The strength of the current framework is that we start with a microscopic model for the polymer chains, then the constitutive equation is a natural outcome from the variational calculation. 
The exact form of the time-evolution equations will depend on the choice of dissipation function [see Eq. (\ref{eq:bigV})], which then determines the strength of the stress-diffusion coupling. 
The choice of $\alpha_1$ and $\alpha_2$ in Eq. (\ref{eq:bigV}) should be based on the system considered. 
However, the derived stress tensor (\ref{eq:sigma_p}) is independent of this choice and is determined by the specific polymer model.

\begin{acknowledgments}
This work was supported by National Key R\&D Program of China (No. 2022YFE0103800) and the National Natural Science Foundation of China (No. 21774004). The KingFa Company is also acknowledged for funding.
\end{acknowledgments}

\bibliography{ve}


\end{document}